\documentclass[a4paper,10pt]{article}
\usepackage[latin1]{inputenc}
\usepackage[T1]{fontenc}
\usepackage[english]{babel}
\usepackage{txfonts}
\usepackage{latexsym}
\usepackage[square]{natbib} 
\bibpunct{(}{)}{;}{a}{}{,}
\begin{document}
\title{Scalar fields properties for flat galactic rotation curves}
\author{Stéphane Fay\\
Laboratoire Univers et Théories, CNRS-FRE 2462\\
Observatoire de Paris, F-92195 Meudon Cedex\\
France\\
\small{Steph.Fay@Wanadoo.fr}}
\maketitle
\abstract{
The whole class of minimally coupled and massive scalar fields which may be responsible for flattening of galactic rotation curves is found. An interesting relation with a class of scalar-tensor theories able to isotropise anisotropic models of Universe is shown. The resulting metric is found and its stability and scalar field properties are tested with respect to the presence of a second scalar field or a small perturbation of the rotation velocity at galactic outer radii.\\\\
}
Published in Astronomy and Astrophysics, 413, p 799, 2004
\section{Introduction}\label{s1}
One of the most fascinating cosmological problems is dark matter one: 99 percent of the Universe energy density would be hidden. A good indication of dark matter presence is given by galactic rotation curves which disagree with Kepler laws \citep{Fre70}. Particularly, spiral galaxy rotation curves seem flattened at large radii. One possible explanation is that they are made of a luminous disk whose density exponentially decreased to adjust to a dark halo whose distribution evolves as $r^{-2}$ \citep{RubThoForRob97}\\
Dark matter nature is unknown today. From WMAP observations, we know that the matter of which we are made represents 4\% of the Universe content, 23\% is made of cold dark matter and 73\% of dark energy. These exotic types of matter could be represented by scalar fields \citep{MatGuzUre99}, which are predicted by unification theories \citep{EllKalOliYok99}. Starting from this assumption, we will study how they could be responsible for the observed flattened rotation curves. Dark matter is not the only way to explain them \citep{BatFlo00}. Hence, Milgrom \citep{Mil83a,Mil83b,Mil83c}, Sanders \citep{San90} and others have proposed to modify newtonian theory(MOND) for galaxies outer radii and in \cite{BatFlo95}, ad hoc magnetic field are considered.

Hence, the physical framework of this paper will be the scalar tensor theories. Unification theories and particularly supersymmetry predict the existence of scalar fields, which thus deserve be taken into account in cosmology. Most of time, only their cosmological consequences are analysed: quintessence phenomenon \citep{Joh01, Sah01}, isotropisation \citep{Fay01} or inflation for instance. However, they could also be present at galactic scale. In this work, we are going to assume that the dynamics of galaxies at outer radii is described by a scalar tensor theory with a dust perfect fluid, neglecting the radiation. The scalar field $\phi$ will be minimally coupled, massive with a Brans-Dicke function representing its coupling with the metric. It is equivalent to consider that, at galactic scale, the gravitational function is a constant but the potential $U$ and the Brans-Dicke coupling function $\omega$ vary with $\phi$.

Concerning the geometrical framework, we will consider a spherical and static metric. These are reasonable assumptions since, generally, a galaxy has a rotation axis around which turn the stars with a velocity much smaller than light speed. We thus neglect dragging effects, justifying a static metric \citep{GuzMat00}. We will be interested by galactic regions where rotation curves flatten and where most of the dark matter should be present, i.e. the galaxies outer radii. Indeed internal regions need few or not dark matter to explain their dynamics.

Our goal will be to study the form of the metric and the scalar field properties explaining why the galactic rotation curves are flattened. A similar work has been done in \cite{MatGuzNum00} and \cite{MatGuz01}. In the first paper, a massive scalar tensor theory is studied with a fixed $\omega$ but an unknown $U$. After having found the metric compatible with flat rotation curves, it is shown that the only potential able to reproduce such a dynamics should have an exponential form, $U=e^{k\phi}$. In the second paper, with the same form for $\omega$, two massive scalar fields and a perfect fluid are considered. One of the potentials is assumed to have an exponential form and similar results are found. In the present paper, we are going to consider a single massive scalar field with a perfect fluid but both $\omega$ and $U$ will be unknown functions of $\phi$. Then we will look for the metric and relations between $\omega$ and $U$ allowing to get the observed flat rotation curves, thus generalising the results of \cite{MatGuzNum00} to a larger class of scalar tensor theories. Moreover, we will test the stability of our results by considering a small perturbation of the rotation velocity or/and an additional scalar field.\\
It is important to note that other types of rotation curves exist, such as decreasing rotation curves found by Casertano and van Gorkom \citep{CasGor91}. Moreover, it seems that bright compact galaxy rotation curves are slightly decreasing whereas low luminosity ones tend to be increasing, indicating that they have more dark matter. However in this work we will only take into account asymptotically flat rotation curves. Indeed, a large number of them seems to be well approximated by an Universal Rotation Curve \citep{PerSalSte96} whose formulation, adapted to spiral galaxies, tends to a constant at late times. It shows how rotation curves depend on galaxy luminosity: increasing or decreasing rotation curves would respectively correspond to low or high galactic luminosity but should tend to a constant at outer radii. It seems to be confirmed by Swaters \citep{Swa99} who has examined a large number of dwarf galaxies and found that their rotation curves flattened over 2 disk scale lengths. MOND theories or the presence of electromagnetic fields can also predict this type of curves. Consequently, although all the rotation curves do not flatten, this type of behaviour is sufficiently observed or predicted to justify a particular attention.

The plane of this paper is the following. In section 2, we look for metric form and scalar field properties allowing to get flattened rotation curves. In section 3, we discuss about these results.
\section{Metric and scalar field mathematical properties}\label{s2}
This section is divided in two parts. In the first one, we consider the presence of a single scalar field. We look for the metric and properties of the unknown functions $\omega$ and $U$ such that the rotation curves be flattened at galaxy outer radii. In the second one, we test our results stability by adding a second scalar field. We will use a static and spherical metric written as:
\begin{equation}\label{metrique}
ds^2=-e^{2\phi}dt^2+e^{2\Lambda}dr^2+r^2(d\theta^2+sin^2\theta d\phi^2)
\end{equation}
$\phi$ and $\Lambda$ being some functions of $r$.	
\subsection{With a single scalar field}\label{s21}
The action for a minimally coupled and massive scalar field with a perfect fluid is given by:
\begin{equation}\label{action1}
S=\int(R-\frac{\omega}{\psi}\psi_{,\mu}\psi^{,\mu}-U+\frac{16\pi}{c^4}L_m)\sqrt{-g}d^4x
\end{equation}
$R$ is the Ricci scalar, $\psi$ the scalar field, $\omega(\psi)$ the Brans-Dicke coupling function and $U(\psi)$ the potential. $L_m$ is the Lagrangian describing a dust perfect fluid whose impulsion-energy tensor writes $T^{\alpha\beta}=\rho u^ \alpha u^\beta$ with $u^\alpha$ the 4-velocity and $\rho$ the density of the dust fluid. We get the field equations and Klein-Gordon equation by varying the action with respect to the metric functions and scalar field:\\
\begin{equation}\label{00}
r^{-2}\left[r(1-e^{-2\Lambda})\right]'=\frac{\omega}{2\psi}\psi'^2e^{-2\Lambda}+1/2U+\rho
\end{equation}
\begin{equation}\label{rr}
-r^{-2}(1-e^{-2\Lambda})+2r^{-1}\phi'e^{-2\Lambda}=\frac{\omega}{2\psi}\psi'^2e^{-2\Lambda}-1/2U
\end{equation}
\begin{equation}\label{thetatheta}
e^{-2\Lambda}(\phi''+\phi'^2+\phi'\Lambda'-\Lambda'/2)=-\frac{\omega}{2\psi}\psi'^2e^{-2\Lambda}-1/2U
\end{equation}
\begin{equation}\label{KG}
e^{-2\Lambda}\psi'^2(\omega\psi^{-1})_\psi+2e^{-2\Lambda}\omega\psi^{-1}\left[(2r^{-1}-\Lambda'+\phi')\psi'+\psi''\right]-U_\psi=0
\end{equation}
A prime stands for a derivative with respect to $r$ and a $\psi$ indice, a derivative with respect to the scalar field. By subtracting equations (\ref{00}-\ref{rr}) and by summing (\ref{rr}-\ref{thetatheta}), it comes:
\begin{equation}\label{eq1}
r(\Lambda'-\phi')-1+e^{2\Lambda}\left[1-1/2r^2(U+\rho)\right]=0
\end{equation}
\begin{equation}\label{eq2}
r^2\Lambda'(2\phi'-1)+2r^2(\phi''+\phi'^2)+4r\phi'+2-2e^{2\Lambda}(1-r^2U)=0
\end{equation}
Then, we derive $U$ and $\rho$ as some functions of $\Lambda$ and $\phi$:
\begin{equation}\label{rho}
\rho=1/2e^{-2\Lambda}r^{-2}\left[-2+2e^{2\Lambda}+r(4-r+2r\phi')\Lambda'+2r^2(\phi'^2+\phi'')\right]
\end{equation}
\begin{equation}\label{U}
U=-1/2e^{-2\Lambda}r^{-2}\left[2-2e^{2\Lambda}+r\phi'(4+2r\phi')+r^2\Lambda'(2\phi'-1)+2r^2\phi''\right]
\end{equation}
We introduce these expressions in (\ref{thetatheta}) to get $\omega$ as a function of $\Lambda$, $\phi$ and $\psi$. Then putting the above forms of $U$, $\rho$ and $\omega$ in Klein-Gordon equation yields:
\begin{equation}\label{eq3}
\Lambda'\left[2(r-2)+r(r-12)\phi'-2r^2\phi'^2\right]-2\phi'\left[e^{2\Lambda}-3+r^2(\phi'^2+\phi'')\right]=0
\end{equation}
which is scalar field independent. Since we are interested by flat rotation curves, we assume that rotation velocity tends to a constant for large $r$. However, rotation curves as seen by an observer at infinity for a spherical symmetry, are given by $V_{rot}=\sqrt{rg_{tt,r}/(2g_{tt})}$, as shown in \cite{MatGuzNum00} where a newtonian interpretation of this last expression is given. It implies that $\phi'\rightarrow V_{rot}^2r^{-1}$ and then $e^{2\phi}\rightarrow r^{2V_{rot}^2}$. To simplify our results below, we define the following constants:
\begin{eqnarray*}
c_1&=&2(1+V_{rot}^2-2V_{rot}^6)\\
c_2&=&V_{rot}^4-1\\
c_3&=&-2(2+6V_{rot}^2+V_{rot}^4)\\
c_4&=&2+V_{rot}^2\\
c_5&=&-2V_{rot}^2(-3-V_{rot}^2+V_{rot}^4)(2+6V_{rot}^2+V_{rot}^4)^{-1}\\
c_6&=&-3-V_{rot}^2+V_{rot}^4\\
c_7&=&-4-12V_{rot}^2-2V_{rot}^4\\
c_8&=&-2-4V_{rot}^2-4V_{rot}^4\\
\end{eqnarray*}
Introducing $\phi'$ in (\ref{eq3}) and integrating, we find for $\Lambda$:
\begin{equation}\label{Lambda}
e^{2\Lambda}=c_6\left[\Lambda_0(\frac{c_4r+c_7}{r})^{c_5}-1\right]^{-1}
\end{equation}
$\Lambda_0$ is a positive integration constant. This last expression is only physically meaning for large $r$ where it tends to the constant $c_6\left[\Lambda_0c_4^{c_5}-1\right]^{-1}$ as $1/r$. This constant must be positive otherwise $e^\Lambda$ is not defined for large $r$ and moreover, numerical integrations seem to show that $\Lambda$ diverges for a finite value of this coordinate. Then, from (\ref{rho}), (\ref{U}) and (\ref{Lambda}), we calculate that asymptotically:
\begin{equation}
\rho=2\left[-1+\Lambda_0\frac{(c_1+c_2r)(c_3r^{-1}+c_4)^{c_5}}{c_3+c_4r}\right](c_6r^2)^{-1}
\end{equation}
\begin{equation}
U=2\{c_2-\frac{(2c_4-3)\Lambda_0\left[c_8+r(c_4-1)\right](c_4+c_3r^{-1})^{c5}}{c_3+c_4r}\}(c_6r^2)^{-1}
\end{equation}
For large $r$, $\rho$ and $U$ vanish as $r^{-2}$. This asymptotical behaviour for the perfect fluid energy density is the same as the nonsingular isothermal profile, one of the most frequent halos. The metric describing the galaxies outer radii where the rotation curves flatten is thus the same as in \cite{MatGuzNum00} whatever $\omega$. Considering a perturbation $\delta(r)$ of $V_{rot}$ does not modify these results as long as $r\delta'\rightarrow 0$.\\
From the form of the metric and since we have left $\omega$ undetermined, we can get for large $r$ a relation between $\omega$ and $U$ such that the rotation curves be flattened. By summing (\ref{00}) and (\ref{rr}) and taking into account asymptotical behaviours for $\rho$ and $U$, we find the following three limits:
\begin{eqnarray}
\omega\psi'^2\psi^{-1}&\rightarrow &4\ell^{-2}r^{-2}\label{ap1}\\
U&\rightarrow &U_1r^{-2}\label{ap2}\\
U'=U_\psi\psi'&\rightarrow &-2U_1r^{-3}\label{ap3}\\\nonumber
\end{eqnarray}
$\ell^{-2}=\left[c_4-3+\frac{c_6}{c^4(\Lambda_0^{-1}c_4^{-c_5}-1)}\right]$ and $U_1$ are some constants. We use (\ref{ap2}) and (\ref{ap3}) for respectively introduce $U$ and replace $\psi'$ in (\ref{ap1}). Then, considering $g$ and $k$ two functions of $r$ and rewriting (\ref{ap1}) and (\ref{ap3}) as respectively $\omega\psi'^2\psi^{-1}\rightarrow g(r)$ and $U_\psi\psi'\rightarrow k(r)$, we have 
$$
\frac{\omega k^2}{\psi U_\psi^2}\rightarrow gr^4r^{-4}
$$
Using (\ref{ap2}) to replace $r^4$ and introduce $U$, it comes
$$
\frac{\omega U^2}{\psi U_\psi^2}\rightarrow U_1^2\frac{g(r)}{k(r)^2}r^{-4}$$
Since here $g(r)=4\ell^{-2}r^{-2}$ and $k(r)=-2U_1r^{-3}$, we find:
\begin{equation}\label{iso}
\frac{\psi U_\psi^2}{\omega U^2}\rightarrow \ell^2\not =0
\end{equation}
For a given form of $U(\psi)$, (\ref{ap2}) defines a unique form for $r(\psi)$. $U(\psi)$ and $r(\psi)$ may then be introduced in (\ref{ap3}), defining a unique form for $\psi'(\psi)$. Then, $r(\psi)$ and $\psi'(\psi)$ may be introduced in (\ref{ap1}), defining a unique form for $\omega(\psi)$. Consequently, for a given $U(\psi)$, (\ref{ap1}-\ref{ap2}) defined a unique $\omega(\psi)$. The same remark applies to (\ref{iso}). Consequently, for a given $U$, the system (\ref{ap1}-\ref{ap3}) or (\ref{iso}) uniquely define $\omega$ and thus the class of scalar tensor-theories responsible for the rotation curves flattening for given potential or Brans-Dicke coupling function. However, only (\ref{ap1}-\ref{ap3}) uniquely define $\psi(r)$.\\
We have shown above that $\rho$ asymptotically behaves as $r^{-2}$. Examining the equation (\ref{00}), we note that the scalar field energy density $\rho_\phi$ shall be written as $\frac{\omega}{2\psi}\psi'^2e^{-2\Lambda}+1/2U$. Knowing the asymptotical limits of each of these terms, we deduce that for large $r$, $\rho\propto \rho_\phi$: the scalar field is quintessent.\\
The limit (\ref{iso}) is doubly important. \underline{Firstly}, in \citep{Fay01,Fay03,FayLum03} it has been shown that a necessary condition for isotropisation of Bianchi models was $\frac{\psi U_\psi^2}{\omega U^2}\rightarrow \ell^2$, $\ell$ being a constant in a close interval depending on the presence of curvature and perfect fluid. Consequently, galactic scalar field properties for large $r$ could match a cosmological scalar field present in the entire Universe which would allow for its isotropisation. \underline{Secondly}, specifying one of the unknown functions $\omega$ or $U$, (\ref{iso}) allow determining in a unique way the other one: this limit gives a necessary and sufficient relation between these two functions such that the galactic rotation curves for outer radii be flattened. It thus generalises the work of \cite{MatGuzNum00} for which $\omega$ was a known function of the scalar field leading to an exponential potential.\\
In the following section, we examine the stability of these results with respect to the presence of a second scalar field.
\subsection{With 2 scalar fields}\label{s22}
When two massive scalar fields are present, the action may take the following form:
\begin{equation}\label{action2}
S=\int(R-\frac{\omega_1}{\psi_1}\psi_{1,\mu}\psi_1^{,\mu}-\frac{\omega_2}{\psi_2}\psi_{2,\mu}\psi_2^{,\mu}-U+\frac{16\pi}{c^4}L_m)\sqrt{-g}d^4x
\end{equation}
The $\psi_i$ are two scalar fields such that $\omega_1=\omega_1(\psi_1)$, $\omega_2=\omega_2(\psi_2)$ and $U=U(\psi_1,\psi_2)$. This form of the action is not the most general one but allows testing the results of the previous section. The field equations are:
\begin{equation}\label{002}
r^{-2}\left[r(1-e^{-2\Lambda})\right]'=\frac{\omega_1}{2\psi_1}\psi_1'^2e^{-2\Lambda}+\frac{\omega_2}{2\psi_2}\psi_2'^2e^{-2\Lambda}+1/2U+\rho
\end{equation}
\begin{equation}\label{rr2}
-r^{-2}(1-e^{-2\Lambda})+2r^{-1}\phi'e^{-2\Lambda}=\frac{\omega_1}{2\psi_1}\psi_1'^2e^{-2\Lambda}+\frac{\omega_2}{2\psi_2}\psi_2'^2e^{-2\Lambda}-1/2U
\end{equation}
\begin{equation}\label{thetatheta2}
e^{-2\Lambda}(\phi''+\phi'^2+\phi'\Lambda'-\Lambda'/2)=-\frac{\omega_1}{2\psi_1}\psi_1'^2e^{-2\Lambda}-\frac{\omega_2}{2\psi_2}\psi_2'^2e^{-2\Lambda}-1/2U
\end{equation}
\begin{equation}\label{KG1}
e^{-2\Lambda}\psi_1'^2(\omega_1\psi_1^{-1})_{\psi_1}+2e^{-2\Lambda}\omega_1\psi_1^{-1}\left[(2r^{-1}-\Lambda'+\phi')\psi_1'+\psi_1''\right]-U_{\psi_1}=0
\end{equation}
\begin{equation}\label{KG2}
e^{-2\Lambda}\psi_2'^2(\omega_2\psi_2^{-1})_{\psi_2}+2e^{-2\Lambda}\omega_2\psi_2^{-1}\left[(2r^{-1}-\Lambda'+\phi')\psi_2'+\psi_2''\right]-U_{\psi_2}=0
\end{equation}
Making the same calculus as in section \ref{s21}, we get an equation similar to (\ref{eq3}), i.e. independent on the scalar fields:
\begin{eqnarray}\label{eq4}
4-4e^{2\Lambda}+4r^3\phi'^3+2r^3\Lambda'^2(2\phi'-1)+r^3\Lambda''-2r\phi'(4-2e^{2\Lambda}+r^2\Lambda'')-\nonumber&\\
4r^2\phi''+2r\Lambda'\mbox{[}6-2r-(r-16)r\phi'+4r^2\phi'^2+r^2\phi''\mbox{]}-2r^3\phi'''=0&\\\nonumber
\end{eqnarray}
Equations for $\rho$ and $U$ are the same as (\ref{rho}) and ($\ref{U}$). Equation (\ref{eq4}) does not depend on $U$, $\omega_1$, $\omega_2$ or the scalar fields forms. Moreover, we always have $\phi'\rightarrow V_{rot}^2r^{-1}$ which asymptotically characterises a flat rotation curve. The solution for $\Lambda$ issued from equation (\ref{eq4}) is thus independent on the scalar fields and the unknown functions $\omega_i$ and $U$. It will be always the same, whatever $U$, $\omega_1$, $\omega_2$ and $\psi_i$. In particular, if we consider the special case where one of the scalar fields is negligible, one have to recover the same asymptotical form for $\Lambda$ as when only one scalar field is present. Hence, the asymptotical solution for equation (\ref{eq4}) should be the same as for (\ref{eq3}): when 2 scalar fields are considered, $\Lambda$ tends to a constant as $r^{-1}$ and $\Lambda'$ vanishes as $r^{-2}$. This is in agreement with \cite{MatGuz01} and implies that $U$ and $\rho$ also vanish as $r^{-2}$. These results are the same if we consider a perturbation $\delta$ for the rotation velocity as long as $\delta'r$ and $\delta''r^2$ are asymptotically vanishing.\\
Anew, we find the following limits allowing to determine if a relation exists between $\omega_1$, $\omega_2$ and $U$ when the rotation curves flatten:
\begin{eqnarray}
\omega_1\psi_1'^2\psi_1^{-1}+\omega_2\psi_2'^2\psi_2^{-1}&\rightarrow &2\ell^{-2}r^{-2}\\
U&\rightarrow &U_1r^{-2}\\
U'=U_{\psi_1}\psi'_1+ U_{\psi_2}\psi'_2&\rightarrow &-2U_1r^{-3}\\\nonumber
\end{eqnarray}
Let us put that $\omega_i\psi_i'^2\psi_i^{-1}\rightarrow g_i$ and $U_{\psi_i}\psi'_i\rightarrow k_i$. Then, $g_1+g_2\rightarrow r^{-2}$, $k_1+k_2\rightarrow r^{-3}$ and we have 
$$
\frac{\omega_i U^2}{\psi_i U_{\psi_i}^{2}}\rightarrow U_1^2\frac{g_i}{k_i^2}r^{-4}
$$
It implies that only one of the $g_i$(or $k_i$) have to tend to $r^{-2}$ (respectively $r^{-3}$), the second one varying as or slower than this last limit. Moreover, equations (\ref{KG1}-\ref{KG2}) may be written as:
$$
(\frac{\omega_i}{\psi_i}\psi_i'^2e^{-2\Lambda})'+2\frac{\omega_i}{\psi_i}\psi_i'^2e^{-2\Lambda}(\frac{2}{r}+\phi')-U_{\psi_i}\phi_i'=0
$$
Hence, since $\Lambda\rightarrow const$ and $\phi'\rightarrow r^{-1}$, if $g_i<r^{-2}$, $k_i$ have to vary slower than $r^{-3}$. We thus distinguish 2 possible behaviours for $g_i$ and $k_i$:
\begin{itemize}
\item case 1: $g_i\rightarrow r^{-2}$ and $k_i\rightarrow r^{-3}$\\
As previously, it comes that $\frac{\psi_1 U_{\psi_1}^{2}}{\omega_1 U^2}$ and $\frac{\psi_2 U_{\psi_2}^{2}}{\omega_2 U^2}$ tend to some constants.
\item case 2: $g_1\rightarrow r^{-2}$, $g_2<<r^{-2}$, $k_1\rightarrow r^{-3}$ et $k_2<<r^{-3}$\\
Consequently, the dynamical effects of $\psi_2$ are asymptotically negligible in the field equations and the metric functions dynamics does not depend on it. We find that $\frac{\omega_1 U^2}{\psi_1 U_{\psi_1}^2}$ tends to a non vanishing constant and $\frac{\omega_2 U^2}{\psi_2 U_{\psi_2}^2}$ diverges or vanishes.\\
\end{itemize}
We conclude that the scalar fields $\psi_i$ which are not asymptotically negligible are such that $\frac{\psi_i U_{\psi_i}^2}{\omega_i U^2}$ tend to some non vanishing constants. Hence, the results of the previous section are not modified by the introduction of a second scalar field. Anew, we note that the scalar fields energy density which shall be $\frac{\omega_1}{2\psi_1}\psi_1'^2e^{-2\Lambda}+\frac{\omega_2}{2\psi_2}\psi_2'^2e^{-2\Lambda}+1/2U$ asymptotically tends to $r^{-2}$ and behaves as the one of the dust fluid. It means that the two (or the dominant) scalar fields are (respectively is) quintessent.
\section{Discussion}\label{s3}
In this work, we have looked for the characteristics of the metric functions and scalar field such that galactic rotation curves flatten for outer radii. For the metric we have got the following result:\\
\\
\emph{Let us consider a minimally coupled and massive scalar field defined by a potential $U$ and a Brans-Dicke coupling function $\omega$ with a spherically static metric. When, at outer radii, the galactic rotation curves flatten, the metric is asymptotically defined by $ds^2=-r^{2V_{rot}^2}dt^2+\Lambda_1dr^2+r^2(d\theta^2+sin^2\theta d\phi^2)$, $\Lambda_1$ being a constant, whatever the form of the potential or Brans-Dicke coupling function.}\\
\\
This result is stable related to a small perturbation $\delta$ of the rotation velocity such that $r\delta'\rightarrow 0$ or if we consider a second scalar field $\phi$ as defined by the Lagrangian (\ref{action2}). In this last case the perturbation must be such that $\delta'r$ and $\delta''r^2$ vanish for large $r$. It is in accordance with the asymptotical form of the metric found in \cite{MatGuz01}  where an exponential potential was found to explain the flattening of the rotation curves. Here, this property is generalized to any functions $\omega$ and potential $U$ with the following characteristics:\\
\\
\emph{Let us consider a minimally coupled and massive scalar field defined by a potential $U$ and a Brans-Dicke coupling function $\omega$ with a spherically static metric. When, at outer radii, the galactic rotation curves flatten, the energy densities of the perfect fluid and scalar field vanish as $r^{-2}$: the scalar field is asymptotically quintessent. Moreover, $\omega$ and $U$ are asymptotically related by the relation $\frac{\psi U_\psi^2}{\omega U^2}\rightarrow \ell^2$, $\ell^2$ being a constant, and U vanishes as $r^{-2}$.}\\
\\
Hence the energy density of the scalar field shows that we are using an isothermal profile which decays in $r^{-2}$ and fit the flat galactic rotation curves quite well and not a Navarro-Frenk-White \citep{NavFreWhi97} profile where the density goes like $r^{-3}$ in the asymptotic region and whose corresponding metric and rotation velocity has been recently determined in \cite{MatNun03}. When 2 scalar fields are present, again their energy density behaves as the one of the perfect fluid. Consequently, at least one of them is quintessent. The presence of quintessent scalar fields in spiral galaxies have been examined in \cite{MatGuz00} where the agreement with the observed rotation curves is shown. Moreover, the scalar fields which are not negligible are such that $\frac{\psi_i U_{\psi_i}^2}{\omega_i U^2}$ tends to a constant, leaving the above last result unchanged for these scalar fields.\\

Let us make some remarks on the quantity $\frac{\psi U_\psi^2}{\omega U^2}$. Firstly, \emph{any necessary condition expressed with the unknown functions\footnote{i.e. $\omega$ and $U$ for the present case.} of a scalar tensor theory such that the metric converge toward a determined form must be invariant with respect to a scalar field transformation}. Indeed, considering $F(U,\omega)$, a necessary condition such that $ds^2$ always tends to a determined form, since a transformation $\psi=T(\Psi)$ of the scalar field keeps the metric, it must be the same for the necessary condition $F(U,\omega)$. One can easily check it is the case for $\frac{\psi U_\psi^2}{\omega U^2}$. Particularly there is a scalar field transformation which allows to rewrite the metric under the form 
$$
S=\int(R-\Psi_{,\mu}\Psi^{,\mu}-U+\frac{16\pi}{c^4}L_m)\sqrt{-g}d^4x
$$
corresponding to the one of \cite{MatGuzNum00} and leading to their results. However, for most of $\omega$ functions, this transformation is not defined or analyticaly workable and thus the results of \cite{MatGuzNum00} can not be arbitrarily applied to any forms of couple $(\omega,U)$ whereas it could be important to keep both $\omega$ and $U$ as depending on the scalar field. As instance, if the potential vanishes, thus mimicing a vanishing cosmological constant, compatibility of the theory with PPN parameters requires that $\omega\rightarrow \infty$ and $\omega_\phi\omega^{-3}\rightarrow 0$ \citep{Nor68,Wag70}, which does not fit with a constant $\omega$ got after field redefinition. Indeed, it recovers the general problem of finding a metric whose dynamics is agreed with the observations and whose potential is in accordance with, as instance, particle physics predictions for the form of the potential: for this, it is necessary to keep the freedom of choosing a form for $\omega(\phi)$. Keeping $\omega$ as an undetermined function of the scalar field thus allows finding the set of all scalar tensor theories able to produce flat rotation curves for \emph{any} forms of $\omega$ and $U$, even when the above scalar field redefinition can not be anatically performed.\\

The part of the results concerning the convergence of $\frac{\psi U_\psi^2}{\omega U^2}$ to a constant seems strangely correlated to isotropisation of homogeneous models which also needs this condition \citep{Fay01,Fay03,FayLum03}. It shows that the properties of a galactic scalar field allowing the flattening of rotation curves could match those of a cosmological scalar field favouring Universe isotropisation.\\
\\
Starting from the form of the potential, the property $\frac{\psi U_\psi^2}{\omega U^2}\rightarrow \ell^2$ allows recovering the Brans-Dicke coupling function such that the rotation curves could flatten and vice-versa. Let us examine some of the most studied potentials. Hence, if we consider an exponential potential $U=e^{k\psi}$, we find that $\omega=k^2\ell^{-2}\psi$, in accordance with the results of \cite{MatGuzNum00} as a particular case of the class of theories we have found, and $\psi\propto \ln r$. If we take $U=\psi^k$, the Brans-Dicke coupling function should be $\omega=k^2\ell^{-2}\psi^{-1}$ and $\psi\propto r^{-2k^{-1}}$. Thus, we see that considering two unknown functions $\omega$ and $U$ instead of a single one leads to an important generalisation of \cite{MatGuzNum00}. Indeed, assuming asymptotically flat rotation curves fix the behaviour of one of the metric function, i.e. $\phi$. Consequently, in \cite{MatGuzNum00}, since $\omega$ is chosen, the potential is uniquely determined whereas in this paper we can find it depends on $\omega$, thus explaining the asymptotical relation between $\omega$ and $U$. It follows that an exponential potential is not the only one allowing to get flat rotation curves but a whole class of scalar tensor theories such that $\frac{\psi U_\psi^2}{\omega U^2}\rightarrow \ell^2$ leads to this property. We remark that a cosmological constant can not explain why the rotation curves flatten since the potential must evolve as $r^{-2}$. This is in agreement with the theories which try to solve the cosmological constant problem by considering it as a variable function rather than a true constant.\\

The interpretation of scalar fields properties found in this work may be done at cosmological or galactic scales. Phenomena giving birth to scalar fields probably have a cosmological nature and are related to particle physics theories. To our knowledge it does not exist any way to generate them by galactic process. Their properties based on rotation curves observations are coherent with their usual cosmological picture i.e. their quintessent nature and their role in the Universe isotropisation. However, if we consider an asymptotically increasing rotation curve, it could mean that $\phi'$ increased faster than $r^{-1}$. It would come from equations (\ref{U}) and (\ref{rho}) that the quantity $U/\rho$ could diverge. If such rotation curves were observed (it is the case but some doubts subsist on the fact that they could asymptotically flatten \citep{PerSalSte96}), it would mean that scalar fields, at least at galactic scale, would not be quintessent. Then the quintessence properties would only be valid for some types of galaxies. In \cite{CasGor91}, decreasing, increasing or flattening rotation curves are studied. Observations seem to show that the first ones appear for bright galaxies and the second ones for faint galaxies. A possible interpretation would be that flat rotation curves would be the outcome of an equivalent mixing between luminous and dark matter, the decreasing or increasing rotation curves resulting of a respectively luminous matter or dark matter domination. This flat rotation curves interpretation is coherent with the scalar field quintessence property found in this paper.\\
\\

To conclude, this work generalises those of \cite{MatGuzNum00} and \cite{MatGuz01}. The metric got in these papers and the quintessent nature of the scalar fields have been generalised for any form of $\omega$. A relation between $U$ and $\omega$ has been found and allows getting easily one of these quantities from the other. It selects the class of scalar tensor theories which could be in agreement with flat rotation curves and shows that an exponential potential is not the only one able to produce such a dynamics. Moreover, we have remarked that this class is also in agreement with Universe isotropisation. The stability of these results with respect to a small perturbation of rotation velocity or the presence of a second scalar field have been tested. A next step would be to consider a non minimally coupled scalar tensor theory, i.e. with a variable gravitational constant or/and a magnetic field that could play a fundamental role at a galactic scale.\\
\section*{Acknowledgements}
I thank Mr Jean-Pierre Luminet for carefull reading of the manuscript and the referee useful comments.
\bibliographystyle{10}

\end{document}